\documentstyle[prb,aps,preprint]{revtex}
\begin{document}

\title{A solvable model for excitonic complexes in \\ 
one dimension } 
\author{Anders J. Markvardsen and Neil F. Johnson}
\address{Department of Physics, Parks Road, Oxford University, Oxford, OX1
3PU,
England}

\maketitle 
\begin{abstract} It is known experimentally that 
stable few-body clusters containing negatively-charged electrons (e) and
positively-charged
holes (h) can exist in low-dimensional semiconductor nanostructures. In
addition to the
familiar exciton (e+h),  three-body `charged excitons' (2e+h and 2h+e) have
also been
observed. Much less is known about the properties of such charged excitons since
three-body problems are generally very difficult to solve, even
numerically. Here we introduce a
simple model, which can be considered as an extended Calogero model, to
calculate 
analytically the energy spectra for both a
charged exciton and a
neutral exciton in a one-dimensional nanostructure, such as a
finite-length quantum wire.
Apart from its physical motivation, the model is of mathematical interest
in that
it can be related to the Heun (or Heine) equation and, as shown explicitly,
highly accurate, closed
form solutions can be obtained. \end{abstract}

\vskip 0.2in

% \narrowtext
\newpage

\noindent{\bf I. Introduction}

The optical properties of low-dimensional semiconductor structures, called
nanostructures, have
attracted much attention in the past few years, both experimentally and
theoretically.
One of the most interesting questions concerns the properties of `excitons'
in such
low-dimensional structures. An exciton (X) is a neutral, two-body complex
formed by the
attractive force between a negatively-charged electron (e) in the
semiconductor conduction band
and a positively-charged hole (h) in the valence band. An exciton therefore
appears to be
somewhat analogous to a hydrogen atom. There are, however, two important
distinctions. First,
the hole and electron masses are typically of the same order of magnitude
and, second, the
low dimensionality of the nanostructure can restrict the electron and hole
motion to
such an extent
that the exciton must be treated as either two- or one-dimensional. 

Recent observations
of anomalies in the optical spectra of quantum wells (i.e. two-dimensional
nanostructures) have been attributed to the formation of negatively-charged
excitons 
\cite{Finkelstein}.  Such complexes can arise when an exciton is created in
the presence of
a low concentration of free electrons; it may then be energetically
favorable for the
exciton to capture one electron to form a negatively-charged exciton, i.e. ${\rm
X}+e\rightarrow {\rm X}^-$. In addition to two-dimensional nanostructures,
it is interesting to
consider the possibility of X$^-$ formation in one-dimensional
nanostructures such as a quantum
wire. The consideration of such low-dimensional systems is particularly
important since the
exciton binding energies in a quantum wire are higher than those in the
quantum well, owing to
the reduced dimensionality; this increased exciton binding energy is
thought to underlie the
recently observed exciton lasing in a quantum wire device \cite{West}. 

This paper uses a simple model to investigate the properties of three-body
complexes such as charged excitons. In particular, we provide closed-form
expressions for the
energy spectra of both a charged exciton X$^-$ and a neutral exciton X in a
finite-length,
one-dimensional quantum wire. Even in one dimension, the three-body problem
(i.e. two electrons
(e) and one hole (h)) with a Coulomb interaction would necessitate a
computationally-intensive
numerical solution.
This is not our goal; instead we wish to demonstrate that
analytically-solvable models can be
introduced to identify trends in the X$^-$ and X energy spectra as a
function of device
parameters. We therefore sacrifice quantitative accuracy concerning a
particular device in
favour of a broader understanding of qualitative behaviour. 

Our model considers an
inverse-square interaction potential between particles (e-h is attractive
and e-e is
repulsive) together with a parabolic confinement potential 
of
arbitrary strength along the wire.  A non-Coulombic form for the
interaction is in fact not
unrealistic
in nanostructures due to the presence of image charges in neighbouring gates
and
electrodes (see, for example, Ref. 3). As will be discussed, 
the energy spectrum of the
neutral exciton X shows the same qualitative behaviour with both 
$\frac{1}{x^{2}}$ (i.e. inverse-square) and $\frac{1}{|x|}$ (i.e. Coulomb)
interactions.
For X$^-$, the three-body  Schrodinger equation is shown to reduce to
the Heun equation. The complete energy spectrum is found in the regime
of physical interest. 
The analysis suggests that the X$^-$ complex 
can have an enhanced stability as compared to
X. $\ \ \ $

\vskip\baselineskip

\noindent{\bf II. Neutral exciton X}

Our model Hamiltonian for the neutral exciton X (i.e. e-h pair) in a 
finite-length one-dimensional quantum wire is given 
by 
\begin{equation} H = -
\frac{\hbar^{2}}{2m^{*}} \left( \frac{\partial^{2}}{\partial x_{e}^{2}} +
\frac{\partial^{2}}{\partial x_{h}^{2}} \right) + \frac{1}{2} m^{*}
\omega_{0}^{2} (x_{e}^{2}+x_{h}^{2})- \frac{q_{eh} \hbar^{2}}
{2m^{*}}\frac{1}{(x_{e}-x_{h})^{2}}\ ,  
\label{neuX}
\end{equation} 
where
$x_{e}$ and $x_{h}$ are the electron and hole coordinates, the
dimensionless
parameter $q_{eh}$ characterizes the electron-hole interaction strength,
and $m^*$
is the effective mass of the electrons and holes (assumed identical). 
The parabolic confinement potential has arbitrary strength and is assumed
to
be the same for both the electron and hole; the confinement parameter
$\omega_0$ can be chosen so as to mimic the effect of a wire of finite
length
$L$ since $L^2\sim \hbar (m^*\omega_0)^{-1}$. If $q_{eh}>\frac{1}{2}$
in Eq. (1) then it is trivial to show that the two particles collapse
toward each other for any finite value of $E$; this can be seen by examining the
behavior of the wave function near $x_e \simeq x_h$. To
avoid this we assume that $0<q_{eh}<{\frac{1}{2}}$.  In terms of the
center-of-mass coordinate $X=\frac{1}{2} (x_{e}+x_{h})$ and the relative
coordinate
$x=x_{e}-x_{h}$, the Hamiltonian becomes
 ${H=H_{c.m.}(X) + H_{rel}(x)}$, where $H_{c.m.}$ is the Hamiltonian for a
single particle in a one-dimensional harmonic potential and
\begin{equation}
H_{rel}= - \frac{\hbar^{2}}{2\mu } \frac{\partial^{2}}{\partial x^{2}} +
\frac{1}{2}
\mu \omega_{0}^{2} x^{2} -\frac{q_{eh}\hbar^{2}}{4\mu}\frac{1}{x^{2}}
\label{relrel} \end{equation} with a reduced mass $\mu =\frac{m^{*}}{2}$.
The
eigenvalues of $H_{c.m.}$ are the one-dimensional harmonic
oscillator levels. The
eigenvalues of $H_{rel}$ are \cite{calogero} 
\begin{equation} E_{rel}(n;q_{eh})=\hbar
\omega_{0} (2n + \frac{3}{2} - \Delta)  \ \ \ \ \ \ \ ;\  n=0,1,2,\ldots
\label{bla} \end{equation} where  
\begin{equation} 
\Delta =\frac{1}{2} -\frac{1}{2}(1-2q_{eh})^{\frac{1}{2}}\ . 
\label{a}
\end{equation}
As the e-h interaction is reduced (i.e. $q_{eh} \rightarrow 0$)  Eq.
(\ref{bla})
becomes
\begin{equation}
E(n;0)=\hbar \omega_{0} (2n+\frac{3}{2}) \ \ \ \ \ \ \ ;\ n=0,1,2,\ldots
\label{mor}
\end{equation}
i.e. the odd energy levels of a harmonic oscillator\cite{calogero}. We shall
refer to the quantity
$\Delta$ as the
electron-hole energy-shift for reasons which are clear by comparing
Eq. (\ref{bla}) and Eq. (\ref{mor}).

The Hamiltonian for the neutral exciton with a Coulomb interaction,
X$_{C}$, 
is the same as 
in Eq. (\ref{neuX}), except that the last term is replaced by the
interaction potential $\frac{1}{|x|}$. We emphasise, however, 
that in the
presence of 
image charges in neighbouring gates etc. \cite{Maksym}, a Coulomb
interaction 
will not necessarily be more realistic than an inverse-square interaction. 
To date, the 
one-dimensional Schrodinger equation for X$_{C}$ has not been completely
solved 
analytically. The Hamiltonian for X$_{C}$ has the form
\begin{equation}
H^C_{rel}= - \frac{\hbar^{2}}{2\mu } \frac{\partial^{2}}{\partial x^{2}} +
\frac{1}{2}
\mu \omega_{0}^{2} x^{2} -\frac{e^2}{\epsilon_{eh}}\frac{1}{x}
\label{Xc}
\end{equation}
where $\epsilon_{eh}$ is the dielectric constant of the system. 

Introducing the
dimensionless coordinate $y$, where $x=by$ with $b=\sqrt{\frac{\hbar}{\mu
\omega_0}}$, the eigenvalue problem of Eq. (\ref{relrel}) reduces to:
$-\psi''(y)+ (y^2 - \frac{q_{eh}}{2}\frac{1}{y^2})\psi
(y)=2E_{rel}/\hbar 
\omega_0 \psi (y)$. Equation (\ref{Xc})
reduces to: 
$-\psi''(y)+ (y^2 - 
\frac{2e^2}{\epsilon_{eh}}\sqrt{\frac{\mu}{\hbar^3\omega_0}}
\frac{1}{y})\psi (y)=2E^C_{rel}/\hbar
\omega_0 \psi (y)$. Typical values
for the device parameters are as follows:
$\epsilon_{eh} = 12$,
$\hbar \omega_0=0.01eV$ and $m^*/m=0.07$. Using these values, we have 
performed numerical calculations
which show that the low-lying energy spectra for X$
_C$ and $X$ can indeed be quantitatively similar, provided an appropriate value 
of the free parameter $q_{eh}$ is chosen. In the $q_{eh}\rightarrow 0$
limit X$_{C}$ yields an {\em
identical} energy spectrum to that given in Eq. (\ref{mor}). 
This indicates that the inverse-square interaction is 'as realistic' as
the bare Coulomb interaction in the context of one-dimensional
models. To
understand why this should be so, one can consider trying to solve the X and
X$_{C}$
problems
using the complete basis set of the one-dimensional oscillator eigenstates
$|n\rangle$ in the interval
$-\infty < x < \infty$, i.e. 
$$ 
\langle x| n \rangle  = \mbox{A}_{n}\mbox{H}_{n}(y){\rm exp}(-y^{2}/2)
$$
where $y = \sqrt{m \omega_{0}/\hbar}\, x$; H$_{n}$ is a Hermite polynomial
and A$_{n}$ is the appropriate normalisation constant. In the complete
basis of kets
$|n\rangle$ we need to evaluate matrix elements of the type \mbox{$\langle
m|H|n\rangle$}, where $H$ is either the Hamiltonian of X or
X$_{C}$. This matrix element includes terms like 
\[
\mbox{A}_{n}\mbox{A}_{m}\int^{\infty}_{-\infty}
\mbox{H}_{n}(y)\mbox{H}_{m}(y)V(y){\rm exp}(-y^{2})dy     
\]
where the potential $V(y)$ is either proportional to $\frac{1}{x^{2}}$ or
$\frac{1}{|x|}$. It is clear that the integral diverges when $m$ and $n$
are
both even. As discussed in Ref. 6, this integral is only finite if we
truncate our Hilbert space so that $m$ and $n$ are odd; 
this finding is hence consistent with the
result of only odd energy levels in Eq. (\ref{mor}). In this sense, both
the $\frac{1}{|x|}$ and
$\frac{1}{x^{2}}$ potentials are `non-penetrable'\cite{kurasov}. 
Physically, this
means that the electrons and hole in the X and X$_{C}$ complexes cannot
interchange their particle ordering along the wire;
the exchange energy is therefore zero. The configuration with
$x>0$ is totally separate from the configuration with $x<0$. Each energy
level of X
and X$_{C}$ is therefore doubly degenerate when the complexes are defined
in the full
interval
$-\infty < x < \infty$. 
Equation (\ref{bla}) represents the complete energy
spectrum of the neutral exciton X with inverse-square interaction.

\vskip\baselineskip
\noindent{\bf III. Charged exciton X$^{-}$} 

Our proposed Hamiltonian for the X$^{-}$ complex may be considered as a
generalisation of the Hamiltonian discussed by Calogero \cite{calogero}. 
Calogero
considered the three-body problem with a harmonic oscillator potential and
inverse-square pair potentials for the case of three identical particles.
Our model Hamiltonian $H$ for two electrons and one hole has the same form,
but
we allow the strength and sign of the interaction between the particles to
be
different. In particular,
$$ H= \sum_{i=e1,e2,h} \left(- \frac{\hbar^{2}}{2m^{*}} 
\frac{\partial^{2}}{\partial x_{i}^{2}} + \frac{1}{2} m^{*} \omega_{0}^{2}
x_{i}^{2} \right)$$
\begin{equation} -
\frac{q_{eh}\hbar^{2}}{2m^{*}}\frac{1}{(x_{e1}-x_{h})^{2}} +
\frac{q_{ee}\hbar^{2}}{2m^{*}}\frac{1}{(x_{e1}-x_{e2})^{2}} -
\frac{q_{eh}\hbar^{2}}{2m^{*}}\frac{1}{(x_{h}-x_{e2})^{2}}
\end{equation} where $x_{e1}$, $x_{e2}$ and $x_{h}$ are the coordinates of
the
two electrons and hole, $q_{ee}$ is the electron-electron interaction
parameter ($q_{ee}>0$); as in Sec. II we restrict the electron-hole
interaction parameter to $0<q_{eh}<{\frac{1}{2}}$. The three-body problem
in
Eq.
(7) is separable\cite{calogero}. The separation involves two
coordinate transformations; first a Jacobi transformation:
\mbox{$X=\frac{1}{3}(x_{e1}+x_{e2}+x_{h})$},
\mbox{$x=2^{-1/2}(x_{e1}-x_{h})$} and
\mbox{$y=6^{-1/2}(x_{e1}+x_{h}-2x_{e2})$} which enables us to rewrite
\mbox{$H=H_{c.m}(X)$} + \mbox{$H_{rel}(x,y)$}. Second, we write $H_{rel}$
in
terms of polar coordinates  $x=r\,sin(\phi -\pi /3)$ and  $y=r\, 
cos(\phi -
\pi /3)$: \begin{equation} H_{rel}(r,\phi )=-\frac{\hbar^{2}}{2m^{*}}
\left(
\frac{\partial^{2}}{\partial r^{2}} + \frac{1}{r} \frac{\partial}{\partial
r}
\right) + \frac{1}{2} m^{*} \omega_{0}^{2} r^{2} + \frac{\hbar^{2}}{2m^{*}}
\frac{M}{r^{2}} , \label{eq:hrel} \end{equation} where $M$ is the following
operator
\begin{equation} M=-\frac{\partial^{2}}{\partial \phi^{2}} + \frac{1}{2}
\left[- \frac{q_{eh}}{sin^{2}(\phi -\pi /3)} + \frac{q_{ee}}{
sin^{2}(\phi
)} -
\frac{q_{eh}}{sin^{2}(\phi +\pi /3)} \right]\  . \end{equation}
Writing an eigenfunction of Eq.
(8) as $\psi (r,\phi )=R(r)\Phi (\phi )$,  the three-body problem
has therefore been reduced to finding the solutions of the
following two ordinary, second-order differential equations
\begin{equation}
H_{rel}R(r)=E_{rel}R(r) \label{eq:difrel} \end{equation} and
\begin{equation}
M\Phi_{l} (\phi )=b_{l}^{2}\Phi_{l} (\phi ) . \label{PHIeq}
\end{equation}
The eigenvalue problem in Eq. (10) is solved in Ref. 4
and the eigenvalues are given by \begin{equation} E_{rel}=\hbar
\omega_{0}(2n
+1 +b_{l}) \ \ \ \ \ \  \ ; n=0,1,2,\ldots \label{RELeigen} \end{equation} 
leaving the nontrivial problem of solving Eq. (\ref{PHIeq}). 
 For a given particle
configuration the angle $\phi$ is limited to a certain interval. The
ordering
of the three particles (i.e. $2{\rm e}+1{\rm h}$) 
is therefore determined by $\phi$. As seen earlier for X, the
`non-penetrable' interaction potentials prevent particle interchange
and therefore make it necessary to treat different particle configurations
separately.
In the interval $\phi \in ]0;\pi [$ the relationship between $\phi$
and the
particle configuration is given by \begin{equation}  \begin{array}{ll} \phi
\in ]0;\pi /3[  & : x_{e2}<x_{e1}<x_{h}  \\

\phi \in ]\pi /3;2\pi /3[ & : x_{e2}<x_{h}<x_{e1}  \\

\phi \in ]2\pi /3;\pi[  & : x_{h}<x_{e2}<x_{e1} . 
\end{array} 
\end{equation} 
Using the trigonometric identity
$sin3\phi =3sin\phi -4sin^{3}\phi$ we can rewrite the
wave function
in Eq. (11) as 
\begin{equation}
\Phi (x)=(x-1)^{\Delta_{ee}/2}(x-\frac{1}{4})^{\Delta_{eh}}y(x)
\end{equation}
where $x= cos^{2}(\phi )$, $\Delta_{eh} = 1-\Delta
=1/2+1/2(1-2q_{eh})^{1/2}$ and
$\Delta_{ee}=1/2+1/2(1+2q_{ee})^{1/2}$. 
Hence we have reduced the problem of solving Eq.
(11) to solving the Heun differential equation
\cite{heun} \begin{equation}
\frac{d^{2}y}{dx^{2}} + \left( \frac{\gamma}{x} +\frac{\delta}{x-1}+ 
\frac{\epsilon}{x-1/4} \right) \frac{dy}{dx} + \frac{\alpha \beta
x-q}{x(x-1)(x-1/4)}y=0 \ , \label{heuneq} \end{equation} with coefficients
\begin{equation} \begin{array}{ll}
 \gamma =1/2 & \alpha =\Delta_{ee}/2 +\Delta_{eh}-1/2 \,   b_{l} \\
 \delta =\Delta_{ee} + 1/2   & \beta =\Delta_{ee}/2 +\Delta_{eh} +1/2\,  
b_{l} \\
 \epsilon =2\Delta_{eh} & 
 q=1/2 (\Delta_{eh})^{2}+(\Delta_{ee}/4)^{2}- (b_{l}/4)^{2} \ 
\end{array} \end{equation} 
and with the five parameters satisfying 
\begin{equation}
\alpha +\beta -\gamma -\delta -\epsilon +1=0 \ .
\label{relation}
\end{equation}
$q$ is the so-called accessory parameter. An introduction
to the general features of Heun's equation is given in Ref. \cite{Arscott}.
In addition, a
recent bibliography containing 300 classified entries related to Heun's
equation are listed in Ref. \cite{Ronveaux}. 
The relation in Eq. (\ref{relation}) ensures that
the four singularities of Eq. (\ref{heuneq}) stay regular. One of
these regular singularities is, however, an elementary singularity (since
$\gamma = 1/2$) which implies that Eq. (\ref{heuneq}) is a special case of
the Heun
equation called the Heine equation \cite{Heine}. To our understanding, Heine's
equation is far less well-known and we will
therefore continue referring to 
Eq. (\ref{heuneq}) as the Heun equation.
Solutions to Eq. (\ref{heuneq}) which are of particular interest are those
which are
analytic in some domain enclosing two singularities. Such
solutions are called `Heun functions' and are denoted by $H$f,
using the notation of Ref. \cite{Arscott}. We
employ this notation in order to distinguish Heun functions $H$f from Heun
polynomials $H$p, which are analytic in an interval containing
{\em three} singularities. Below, Heun functions which are analytic in the
interval $x
\in [0;1/4]$ or $\phi \in [\pi /3;2\pi /3]$ will be studied in more
detail. An eigenfunction of Eq.(\ref{heuneq}) defined in the region
$\phi \in [\pi /3;2\pi /3]$ can be written, using $cos^2\phi =1 -
sin^2\phi$, as
\begin{equation}
\Phi_l(\phi )=(sin\phi )^{\Delta_{ee}}(sin^2\phi
-3/4)^{\Delta_{eh}} H\mbox{f}(1/4,q;\alpha , \beta ,\gamma ,\delta
;cos^2\phi )\ \ .
\label{PHIfun}
\end{equation}
Heun functions may be found by the power-series method or by the method of
hypergeometric function series \cite{Arscott}. However for both
methods the coefficients in the series have to satisfy a three-term
recursion relation. It might be possible to use the method of augmentet
convergence 
to extract Heun functions from such relations, but in most cases only
a numerical procedure is possible \cite{Slavyanov}.
Although we cannot in general solve the eigenvalue problem of Eq. (15)
analytically, we will now 
obtain
a set of highly accurate, approximate solutions for the region 
$\phi \in [\pi /3;2\pi /3]$; these solutions are essentially exact
for a large interval of the ratio $q_{ee}/q_{eh}$, 
including the range of physical interest.
Employing the 
following trigonometric formula \begin{equation} -\frac{f}{
sin^{2}(\phi - \pi
/3)} + \frac{g}{sin^{2}(\phi )} - \frac{f}{sin^{2}(\phi + \pi
/3)}=\frac{f+g}{sin^{2}(\phi )} -\frac{9f}{sin^{2}(3\phi )}\ , 
\end{equation}              where $f$ and $g$ are arbitrary functions
\cite{formula}, the operator $M$ in Eq. (\ref{PHIeq})
 reduces to \begin{equation} M=-\frac{\partial^{2}}{\partial
\phi^{2}} + \frac{q_{eh}}{2} \left[ \frac{1 +\kappa}{sin^{2}\phi}
-\frac{9}{sin^{2}3\phi} \right] \end{equation} where
$\kappa=q_{ee}/q_{eh}$. 
Consider the angular dependence of the total interaction potential of the
 X$^{-}$ complex: $V(\phi )\equiv (1+\kappa )/sin^{2}\phi -
9/sin^{2}3\phi $. Here $V(\phi )$ is periodic in $\phi$, $V(\phi
)=V(\phi +\pi
)$, hence the physics of the three-body problem is contained within a
$\phi$-interval of $\pi$, e.g. $\phi \in ]0;\pi[$. Figure 1 shows $V(\phi
)$
for the case $\kappa =1$. The asymptotic behavior of $V(\phi )$ is
easily understood: at $\phi =0$ the repulsion between the two electrons
causes
a positive singularity, at $\phi = \pi /3$ the attraction between the
electron
at $x_{e1}$ and the hole at $x_{h}$ causes
a negative singularity, etc.  For the particle configuration where the hole
is
between the two electrons (in Fig. 1: $\phi \in ]\pi /3;2\pi /3[$ ), the
potential does not have a repulsive divergence and therefore corresponds
to the most stable three-body configuration. A very good
approximation to $V(\phi )$ in this interval ($\phi \in ]\pi /3;2\pi /3[$)
for
moderate values of $\kappa$ is \begin{equation} V_{app}(\phi ) = 1 +
\kappa - \frac{9}{sin^{2}3\phi}  \ \ . \label{app} \end{equation} 
This is illustrated in
Fig. 2. Evidently the approximation becomes more exact for smaller
$\kappa$ and, as will become clearer later, exact in the limit
$q_{ee}\rightarrow -q_{eh}$. We
shall restrict $\kappa$ to the interval
 $\kappa \in [0;20[$; $V_{app}(\phi )$ is now an excellent approximation to
$V(\phi )$. The quantity $\kappa$ is the ratio of the strength of the 
electron-electron interaction to the strength of the electron-hole
interaction, and for all practical electronic devices we expect this ratio
to be less than 20.
Using this  approximation, Eq. (\ref{PHIeq}) becomes  $M_{app}\Phi_{l}
(\phi )=b_{l}^{2}\Phi_{l} (\phi )$ where $M_{app}=-\partial^{2}/\partial
\phi^{2} + \frac{q_{eh}}{2} V_{app}$. The task of solving this 
differential equation can now be
transformed into the problem of solving a hypergeometric equation. The
solutions are
\begin{equation}
\Phi_l(\phi )=(sin3\phi)^{\Delta_{eh}} {_2F_1}(a,b;c;cos^23\phi )
\label{PHIappfun}
\end{equation}
where 
\begin{equation}
c=1/2 \ , \ a=1/2(\Delta_{eh}-b_l')\ ,\ b=1/2(\Delta_{eh}+b_l') \ 
\end{equation}
and $b_l'^2=1/9 \, b_l^2-1/18(q_{eh}+q_{ee})$. The
exact eigenvalues are found to be \begin{equation} b_{l}=[9(l+1-\Delta
)^{2} +
\frac{1}{2}(q_{ee}+q_{eh})]^{1/2} \ \ \ \ \ \ ; \  l=0,1, \ldots  . 
\label{PHIeigen}
\end{equation}
Combining Eq. (\ref{PHIeigen}) and Eq. (\ref{RELeigen}) gives the 
complete energy
spectrum of the X$^{-}$ complex for the particle
configuration where the hole is  placed between the two electrons and 
$\kappa
\in [0;20[$; in particular, 
 \begin{equation}
E_{rel}(n,l;q_{eh},q_{ee})=\hbar \omega_{0}\left( 2n
+1+[9(l+1-\Delta )^{2}+\frac{1}{2}(q_{ee}+q_{eh})]^{\frac{1}{2}}\right) \ .
\label{comthree}
\end{equation}  
In the Calogero-model limit where $q_{ee}\rightarrow -q_{eh}$, the energy
spectrum in
Eq. (\ref{comthree}) reduces to the energy spectrum found by Calogero for three
identical particles\cite{calogero}. By comparing Eq. (\ref{PHIappfun})
and Eq. (\ref{PHIfun}) 
the approximation corresponding to Eq. (\ref{app})
can be expressed as
\begin{eqnarray}
\Phi_l(\phi ) & = & (sin\phi )^{\Delta_{ee}}(sin^2\phi
-3/4)^{\Delta_{eh}} H\mbox{f}(1/4,q;\alpha , \beta ,\gamma ,\delta
;cos^2\phi ) \nonumber \\ & \simeq & (sin3\phi)^{\Delta_{eh}}
{_2F_1}(a,b;c;cos^23\phi )
\label{appfunc}
\end{eqnarray}
In the Calogero-model limit, Eq.(\ref{appfunc}) becomes exact and
$Hf=(-4)^{\Delta_{eh}} {_2F_1}$.

There exist two distinct particle configurations in which the hole is
positioned
between the two electrons. These two configurations represent
different physically-accessible systems, because of the `non-penetrable'
property of
the interaction potential as discussed in Sec. II. If X$^{-}$ is defined
for the
full range of these two particle configurations, then each level in
Eq. (\ref{comthree}) is doubly-degenerate. By forming suitable linear
combinations, the subspace corresponding to each eigenvalue may be spanned
by an
antisymmetric and symmetric wave function with respect to interchange
of the two electrons. 

\vskip\baselineskip

\noindent{\bf  IV. \, Comparison of energies of X and X$^{-}$}

We have derived the energy spectra for X and X$^{-}$. One possibility might
be
to compare their relative stabilities by turning off
 the confinement potential and calculating the binding energies
of X and X$^-$. However, a finite 
confinement potential is needed within the present model 
in order to produce discrete energy levels for the complexes; 
when the confinement interaction is turned off ($\omega_{0}\rightarrow 0$) 
a continuous spectrum\cite{continuous} is obtained for both X and X$^-$. 
This suggests than X and X$^-$ are not exciton complexes in the usual sense,
since their existence depends on the presence of a confinement
potential. Keeping the confinement potential finite, the ground state energies
of X and X$^-$ in the non-interacting limit ($q_{ee},q_{eh}\rightarrow
0$) are given by ($E_{rel}+E_{c.m}$) which yields $2\hbar \omega_0$ and $9/2
\hbar \omega_0$ respectively; these energies are identical to the 
ground state energies for two and three
spinless fermions in a harmonic potential well. We emphasize that this
particular non-interacting limit is reached as a consequence of the model being
strictly one-dimensional and containing
a singular potential.

Keeping the confinement potential finite, we now investigate the 
{\em changes} in energy of X and X$^-$ as the
 two-body interaction is varied. We introduce a quantity which we refer to
as the
 `interaction energy' $E_{int}$, defined as the energy obtained 
by subtracting the total energy with vanishingly small interactions 
(i.e. $q_{ee},q_{eh}\rightarrow 0$)
 from the total energy with finite
interactions. For X and X$^-$ in their respective ground-states, we obtain
\begin{equation}
E_{int}^{\mbox{{\scriptsize X}}}(q_{eh}) =
E_{rel}(0;q_{eh})-E_{rel}(0;0)=-\hbar \omega_{0}\Delta \
\end{equation}
and
\begin{equation}
E_{int}^{\mbox{{\scriptsize X$^{-}$}}}(q_{ee},q_{eh})=
E_{rel}(0,0;q_{ee},q_{eh})-E_{rel}(0,0;0,0)=
\hbar \omega_{0}
\left ([9(1-\Delta)^{2} + \frac{1}{2}(q_{ee}+q_{eh}) ]^{1/2} -3\right) \ .
\end{equation} 
As expected, both $E_{int}^{\mbox{{\scriptsize X}}}$ and
$E_{int}^{\mbox{{\scriptsize X$^{-}$}}}$ 
become increasingly negative with increasing $q_{eh}$, i.e. 
the ground-state energy decreases as the electron-hole interaction $q_{eh}$
increases. 
In addition
$E_{int}^{\mbox{{\scriptsize X$^{-}$}}}$ increases as 
the electron-electron interaction $q_{ee}$ increases. 
We therefore interpret 
the quantities $E_{int}^{\mbox{{\scriptsize X}}}$ and
$E_{int}^{\mbox{{\scriptsize
X$^{-}$}}}$ as indicative of the binding strength of X and X$^-$
respectively. 
As a consequence, the relative stability of X$^-$ and
X  is then 
effectively represented by the `relative binding strength'
\begin{equation}
\Delta E_{int}(q_{ee},q_{eh})=E_{int}^{\mbox{{\scriptsize X$^{-}$}}}
- E_{int}^{\mbox{{\scriptsize X}}}=\hbar \omega_{0}
\left( [9(1-\Delta )^{2} + \frac{1}{2}(q_{ee}+q_{eh}) ]^{1/2} -3
+\Delta \right)\  . 
\label{dEint}
\end{equation}  
If $\Delta
E_{int} <0$, this would suggest that X$^-$ is more strongly bound than X; 
if $\Delta E_{int}>0$, the reverse is true. The cross-over
occurs when $\Delta E_{int}=0$, i.e. when\begin{equation}
q_{ee}=8 \, \Delta +7 \, q_{eh} \ . 
\label{equilib}
\end{equation} 
We label the $\kappa$ value for which Eq. (\ref{equilib}) is satisfied 
to be $\kappa_{eq}$. When $\kappa<\kappa_{eq}$ for a given $q_{eh}$, then
$\Delta
E_{int}<0$; when $\kappa>\kappa_{eq}$ the reverse is true.
This is illustrated in the inset of Fig. 3 where
$\Delta E$ is plotted as a function of $\kappa$ for the case
$q_{eh}=0.25$. 
The main part of Fig. 3  shows $\kappa_{eq}$ as a
function of $q_{eh}$. Two features are interesting. First, $\kappa_{eq}$
stays within a reasonably narrow interval $11<\kappa_{eq}<15$ for all
 values of $q_{eh}$ within the model. In the range 
 $0<q_{eh}<0.4$, $\kappa_{eq}$ increases only slightly on an
absolute scale. Second, the
minimum value for $\kappa_{eq}$ is given by $\kappa_{eq}=11$. 
Therefore if $\kappa$ is smaller than 11, as is the case for typical
experimental devices, 
this would suggest that X$^-$ is more strongly bound than
X\@. 

\vskip\baselineskip
\noindent{\bf V. Conclusions}

In summary, we studied a new model for a class of three-body problems; 
highly accurate,
closed form solutions for this model were obtained. 
The model was used to study the
exciton complex X$^-$ relative to the neutral exciton X. 
The analysis suggest that X$^-$ might be more
strongly bound than X for typical one-dimensional devices. 

Finally we note that the present results for
X$^-$ apply equally well to X$^+$ (i.e. one electron plus two holes).  
Future
work will use the same model to examine more exotic excitonic
complexes containing $n_e$ electrons and $n_h$ holes, i.e.  X$^{\Delta n \
-}$
where $\Delta n=n_e-n_h$. 

\vskip\baselineskip
\vskip\baselineskip
\noindent{\bf Acknowledgements}

We acknowledge useful discussions with C. Markvardsen, A. Svane, 
G. Bruun and J. Chapman.

\newpage \centerline{\bf Figure Captions}

\bigskip

\noindent Figure 1. 

The potential function $V(\phi )$
for the case $\kappa =1$. Also shown schematically are the
corresponding electron-hole configurations corresponding to the various
$\phi$-regions. 

\bigskip

\noindent Figure 2. 

Comparison of the approximate potential $V_{app}(\phi )$ and the exact
potential
$V(\phi )$ for various values of $\kappa$. For the case $\kappa=1$, 
$V(\phi )$ and
$V_{app}(\phi )$ are essentially identical.
\bigskip

\noindent Figure 3. 

The ratio of the electron-electron interaction to the electron-hole
interaction
at the stability cross-over point, $\kappa_{eq}$, as a function of the
electron-hole
interaction
$q_{eh}$. The inset
shows $\Delta E_{int}$ as a function of $\kappa$ for the case
$q_{eh}=0.25$.


\begin{thebibliography}{99}
\bibitem{continuous} For X, the solution to Eq. (\ref{relrel}) with
$\omega_{0} = 0$ is given by $\psi (x) = x^{1/2}J_{1/2-\Delta}(kx)$ and
$E_{rel} = \frac{\hbar^2}{2\mu}k^2$ ($0\leq k < \infty $). Here
$J_{1/2-\Delta}$ is a Bessel function\cite{calogero}. Parameter $\Delta$ is given in
Eq. (\ref{a}).
\end{thebibliography}
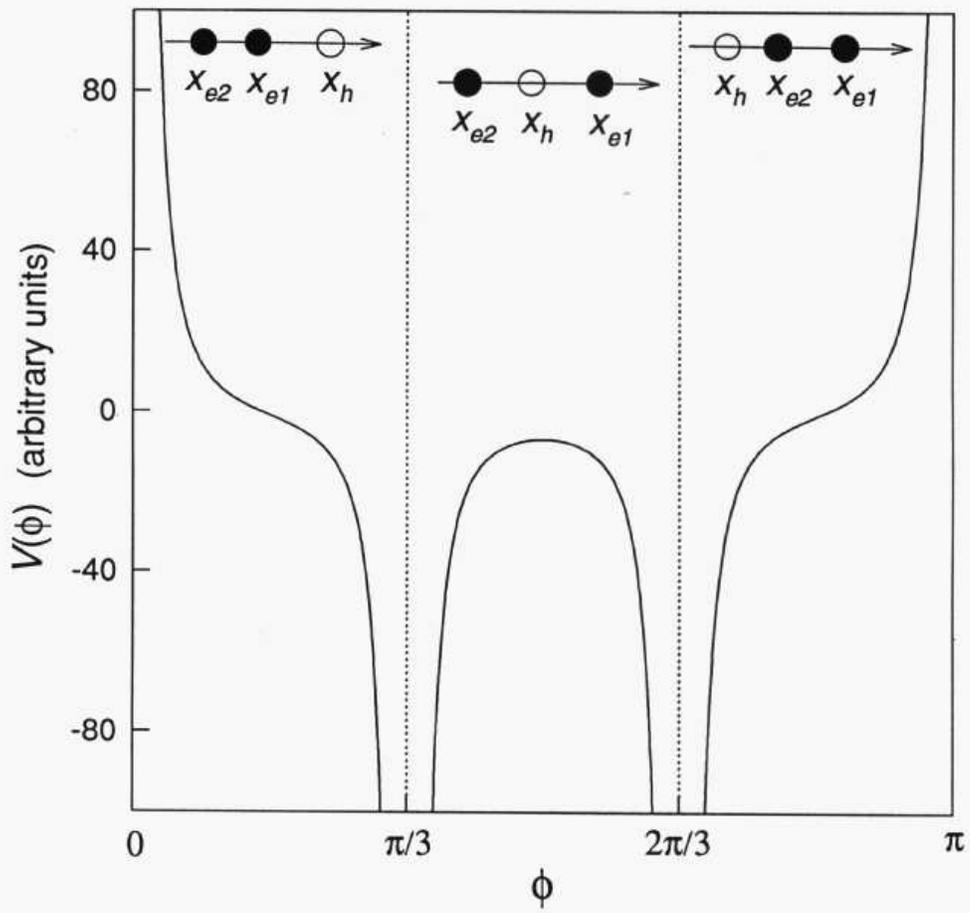
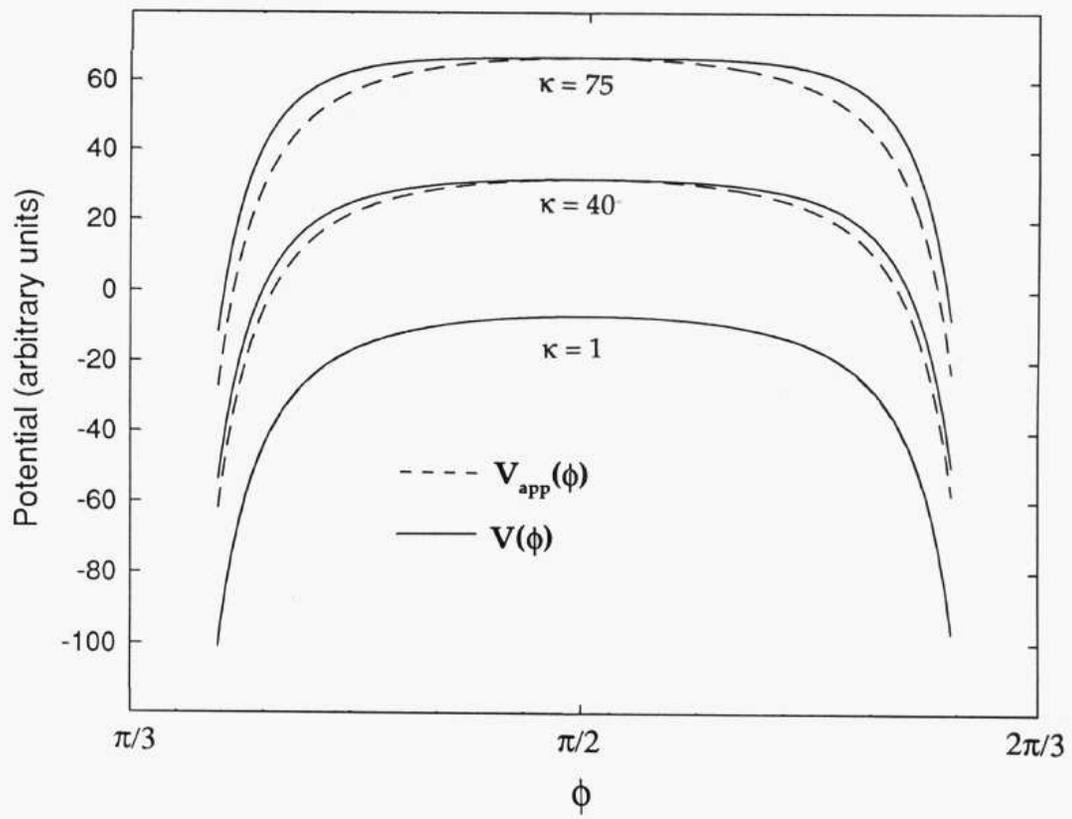
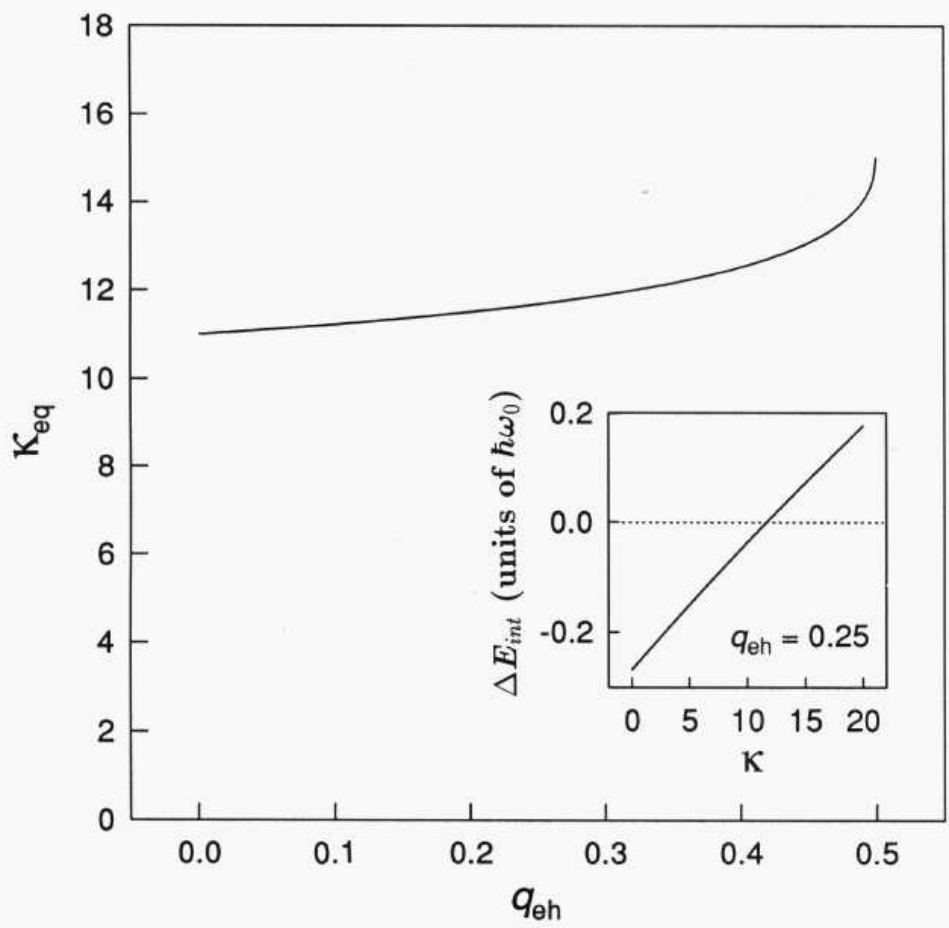
\end{document}